%% file: Bosons_on_ring_arXiv.2.tex
\documentclass[aps,prl,10pt,twocolumn,superscriptaddress]{revtex4-1}

\usepackage{times}
\usepackage{changes}

\usepackage{graphicx}
\usepackage{amssymb}
\usepackage{epstopdf}
\usepackage{amsmath}
\usepackage{color}
\usepackage{hyperref}
\usepackage{bbold}
\usepackage{mathtools}

\usepackage[acronym]{glossaries}
\usepackage{physics}
\makeatletter

\@ifundefined{textcolor}{}
{
 \definecolor{BLACK}{gray}{0}
 \definecolor{WHITE}{gray}{1}
 \definecolor{RED}{rgb}{1,0,0}
 \definecolor{GREEN}{rgb}{0,1,0}
 \definecolor{BLUE}{rgb}{0,0,1}
 \definecolor{CYAN}{cmyk}{1,0,0,0}
 \definecolor{MAGENTA}{cmyk}{0,1,0,0}
 \definecolor{YELLOW}{cmyk}{0,0,1,0}
}

\graphicspath{{Images/}}

\hypersetup{
	pdfpagemode={UseOutlines},
	bookmarksopen,
	pdfstartview={FitH},
	colorlinks,
	linkcolor={black},
	citecolor={blue},
	urlcolor={blue}}
	
\usepackage{hyperref}
\hypersetup{
colorlinks=true,
linkcolor=black,           
citecolor=blue,           
filecolor=magenta,        
urlcolor=cyan
}

\newcommand{\mymod}[1]{\ (\mathrm{mod}\ #1)}

\makeatother

\loadglsentries{Bosons_on_ring_gls}

\begin{document}
\title{Period-$n$ discrete time crystals and quasicrystals with ultracold bosons}

\author{Andrea Pizzi}
\affiliation{Cavendish Laboratory, University of Cambridge, Cambridge CB3 0HE, United Kingdom}
\author{Johannes Knolle}
\affiliation{Blackett Laboratory, Imperial College London, London SW7 2AZ, United Kingdom}
\author{Andreas Nunnenkamp}
\affiliation{Cavendish Laboratory, University of Cambridge, Cambridge CB3 0HE, United Kingdom}

\begin{abstract}
	We investigate the out-of-equilibrium properties of a system of interacting bosons in a ring lattice. We present a Floquet driving that induces clockwise (counterclockwise) circulation of the particles among the odd (even) sites of the ring which can be mapped to a fully connected model of clocks of two counter-rotating species. The clock-like motion of the particles is at the core of a period-$n$ discrete time crystal where $L=2n$ is the number of lattice sites. In presence of a "staircase-like" on-site potential, we report the emergence of a second characteristic timescale in addition to the period $n$-tupling. This new timescale depends on the microscopic parameters of the Hamiltonian and is incommensurate with the Floquet period, underpinning a dynamical phase we call 'time quasicrystal'. The rich dynamical phase diagram also features a thermal phase and an oscillatory phase, all of which we investigate and characterize. Our simple, yet rich model can be realized with state-of-the-art ultracold atoms experiments.
\end{abstract}

\maketitle

\textit{Introduction.---}
Symmetries pervade most fields of modern physics, ranging from nuclear physics to relativity and condensed matter physics. The spontaneous breaking of symmetries is the essential mechanism at the core of equilibrium phase transitions.
Quite surprisingly, the possibility of systems breaking time-translational symmetry has been put forward only recently by Wilczek \cite{shapere2012classical, wilczek2012quantum} who dubbed them 'time crystals' and triggered an extraordinary amount of excitement \cite{sacha2017time, moessner2017equilibration, else2019discrete}.
Since time crystals have been shown to be impossible in quantum ground states \cite{bruno2013impossibility}, research has focussed on out-of-equilibrium conditions.
Particularly successful has been the setting of Floquet systems, characterized by a time-periodic Hamiltonian $H(t) = H(t+T)$, for which the notion of a \gls{DTC} has been introduced in Refs.~\cite{sacha2015modeling, else2016floquet, von2016phase, khemani2016phase}.
A \gls{DTC} is a system that, for some physical observable $O$ with expectation value $f(t) = \expval{O}{\psi(t)}$, in the thermodynamic limit and for a set of initial conditions $\ket{\psi(0)}$, features three properties \cite{russomanno2017floquet}:
(I) discrete time-translational symmetry breaking: $f(t) \neq f(t+T)$;
(II) rigid subharmonic response: without fine-tuning $f(t)$ shows oscillations with a period $nT$, i.e.~period-$n$-tupling, that is the dynamics features a characteristic frequency $\frac{2\pi}{n}$ with $n$ integer $\ge 2$;
(III) persistence: the subharmonic oscillations extend up to infinite time.

Among the plethora of proposed \glspl{DTC}, most feature period-doubling $n=2$ \cite{sacha2015modeling, else2016floquet, von2016phase, khemani2016phase, moessner2017equilibration, ho2017critical, else2017prethermal, iemini2018boundary, zhu2019dicke, russomanno2017floquet, gong2018discrete, yao2017discrete, zhang2017observation, choi2017observation, rovny2018observation, smits2018observation}.
The possibility of a "period-$n$ \gls{DTC}" ($n>2$), which has been discussed for $n$-hands clock models \cite{sreejith2016parafermion, surace2018floquet}, still lacks a physical implementation \cite{giergiel2018time}.
On the other hand, investigations of \glspl{DTC} for ultracold bosons have for the most part remained restricted to the context of a vibrating mirror in presence of a gravitational field \cite{huang2018clean, sacha2015modeling, sacha2015anderson, giergiel2018discrete, giergiel2018time}.
In light of the terrific experimental progress and control with cold atoms in optical lattices \cite{bloch2008many, smits2018observation}, proposals based on this platform would be very desirable.

\begin{figure}[b]
	\begin{center}
		\includegraphics[width=\linewidth]{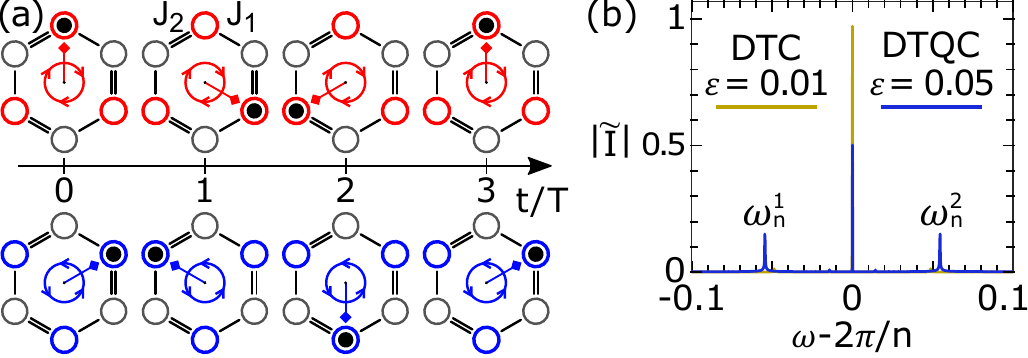}\\
	\end{center}
	\vskip -0.5cm \protect
	\caption{
		\textbf{Two entwined, counter-rotating clocks on a ring lattice.} As a concrete example we show the $n=3$ case.
		(a) Our Floquet driving scheme  alternates (i) hopping between sites $2j$ and $2j-1$ ($J_1$, single line), (ii) hopping between sites $2j$ and $2j+1$ ($J_2$, double line), and (iii) on-site interaction and on-site potential, see Eq.~(\ref{eq: H ternary}).
		For $J_1 = J_2 = \frac{\pi}{2}$ (i.e.~$\epsilon = 0$) the particles in the odd (red) and even (blue) sites move clockwise and counterclockwise, respectively.
		(b) Fourier transform of the generalized imbalance $\tilde{\mathcal{I}}$ in Eq.~\eqref{I_til}. In presence of interaction ($U = 0.05$) and a staircase-like potential ($\delta = 0.96$) we observe a \gls{DTC} (for $\epsilon = 0.01$) and a \gls{DTQC} (for $\epsilon = 0.05$). The former is characterized by a single sharp peak locked at frequency $\frac{2\pi}{n}$, whereas the latter \textit{also} features sharp peaks at two frequencies $\omega_n^{1,2}$ which are incommensurate with the driving frequency.}
	\label{fig: Bosons on ring L=6 scheme}
\end{figure}

In this letter we study the dynamical properties of bosons in a ring lattice. For a simple Floquet driving protocol of nearest-neighbor hopping and local interaction, we find period-$n$ \glspl{DTC} ($n \ge 2$), thermal, and oscillatory phases. Surprisingly, in the presence of a staircase-like on-site potential, a new dynamical phase emerges characterized by three incommensurate frequencies (Fig.~\ref{fig: Bosons on ring L=6 scheme}). The dynamics is strictly aperiodic, but long-time ordered: a \gls{DTQC} \cite{lifshitz2003quasicrystals, lifshitz2011symmetry}.

The idea of time quasi-crystallinity has recently appeared in the literature with various connotations and in various contexts such as dissipative classical systems \cite{flicker2018time}, finite-size systems \cite{huang2018symmetry}, and quasiperiodically driven systems \cite{dumitrescu2018logarithmically, peng2018time, zhao2019}.
The term 'time quasicrystal' has also been adopted in an experiment with magnons in which one of the system's characteristic frequencies was incommensurate with the driving frequency \cite{autti2018observation} and in a model where a periodic repetition in time of a (possibly large, but finite) Fibonacci word was observed \cite{giergiel2018discrete}.

In striking contrast to the cases above, we consider a quantum, macroscopic, and periodically driven system and we find a \gls{DTQC} phase that features a characteristic subharmonic frequency $\frac{2\pi}{n}$ \textit{and} two other, generally incommensurate, intrinsic frequencies $\omega_n^{1,2}$. The first frequency is locked to a submultiple of the driving frequency, breaking discrete time-translational symmetry and robust to perturbations, similar to a standard \gls{DTC}, whereas the latter two frequencies depend on the microscopic parameters.

The remainder of this letter is organized as follows. First, we introduce our model and study its solvable limits. We then propose a set of suitable diagnostic dynamical order parameters to characterize the system. We study the dynamics first in absence of an on-site potential and find a period-$n$ \gls{DTC}, a thermal, and an oscillatory phase.
In the presence of a staircase-like on-site potential, we additionally find a \gls{DTQC} phase. We demonstrate its rigidity, and find analytical expressions for the incommensurate frequencies $\frac{2\pi}{n}$ and $\omega_n^{1,2}$. Finally, we summarize our results and outline possible experimental implementations using state-of-the-art ultracold atom setups.

\textit{Model and integrable limits.---}
We consider a system of $N$ bosons on a one-dimensional ring lattice with $L=2n$ sites governed by the Floquet Hamiltonian with period $T = 1$ \cite{mizuta2018spatial}
\begin{equation}
H = 
\begin{cases}
-3 J_1\sum_{j = 1}^{n} (a_{2j}^\dagger a_{2j-1} + h.c.), \quad & 0 < t < \frac{1}{3} \\
-3 J_2\sum_{j = 1}^{n} (a_{2j}^\dagger a_{2j+1} + h.c.), \quad &\frac{1}{3} < t < \frac{2}{3} \\
3 \sum_{j = 1}^{2n} \left[ \frac{U}{2N}n_{j}(n_{j}-1) + h_{j} n_{j} \right], \quad & \frac{2}{3} < t < 1 \\
\end{cases}
\label{eq: H ternary}
\end{equation}
where we set $J_1 = J_2 = \frac{\pi}{2} + \epsilon$ and $h_{2j-1} = h_{2j} = \frac{2\pi}{n}j \delta$, that is a staircase-like on-site potential. The ring structure implicitly corresponds to periodic boundary conditions $a_{j} = a_{j+L}$. The Floquet Hamiltonian \eqref{eq: H ternary} alternates hopping between odd nearest-neighbor links, hopping between even nearest-neighbor links, and on-site potential and two-body interaction. In the following $t = 0,1,2,\dots$ denotes stroboscopic times.

We gain some intuition in the dynamical properties of the system by solving the Heisenberg equation of motion $\frac{d a_j}{d t} = i [H(t), a_j(t)]$ ($\hbar = 1$) in the two cases $\epsilon = 0$ and $U=0$. The case $\epsilon = 0$ is at the core of a period-$n$ \gls{DTC} and is schematically illustrated in Fig.~\ref{fig: Bosons on ring L=6 scheme}. Consider an initial product state in the site basis with $n_j$ bosons in an odd (even) site $j$. During the first part of the driving these particles move to site $j+1$ ($j-1$), during the second they move to site $j+2$ ($j-2$), and during the third the number of particles in each site is conserved (irrespective of $U$ and $\{h_j\}$). We thus find $n_{j}(t) = n_{j \pm 2t}(0)$ with $+$ and $-$ for even and odd sites $j$, respectively. The bosons in the ring lattice can be shown to be equivalent to a fully connected model of clocks with $n$-hands and of two species: clockwise and counterclockwise rotating (see SM).
The clocks tick at every Floquet period, so that after a time $t=n$ the system returns to its initial condition, a mechanism at the core of a period-$n$ \gls{DTC}.

In the non-interacting limit ($U=0$), the Heisenberg equation of motion is linear and easily solved as $\vec{a}(t+1) = F \vec{a}(t)$, where $\vec{a} = (a_1, a_2 \dots, a_{2n})^T$ and where $F$ is a $2n \times 2n$ dimensional matrix. The dynamics is thus characterized by oscillations at the $2n$ frequencies corresponding to the phases of the $2n$ eigenvalues $\{\lambda_j\}$ of $F$. For instance, for $\epsilon = 0$ we find doubly degenerate eigenvalues of the form $\lambda_j^{1,2} = e^{i\phi + i \frac{2\pi}{n}j}$ (with $\phi$ some non-relevant phase), which correctly signals the period-$n$ \gls{DTC} with clock-like clockwise (counterclockwise) rotation of particles.

\textit{Dynamical order parameters.---}
We are now interested in the dynamics away from the solvable limits $\epsilon = 0$ and $U = 0$. To this end, we solve a semiclassical Gross-Pitaevskii equation of motion, which is obtained from the Heisenberg equation upon replacing the bosonic operators with $c$-numbers $\vec{a} \rightarrow \sqrt{N} \vec{\psi}$. In the thermodynamic limit of macroscopic occupation ($N \rightarrow \infty$) for a fixed finite ring size $L$ \cite{sacha2017time}, considering a symmetry broken initial state with all the particles in site $j=1$
(i.e.~$\psi_j(0) = \delta_{j,1}$), the semiclassical dynamics can either be chaotic, signaling quantum thermalization \cite{cosme2014thermalization} or not, in which case we expect it to become exact \cite{polkovnikov2003quantum}. We indeed confirm this for our model explicitly using exact diagonalization and finite-size scaling (see SM).

It is now crucial to introduce suitable dynamical order parameters. To track the clock-like circulation of the particles we introduce a generalized imbalance on the odd sites
\begin{equation}
\mathcal{I}(t) = \sum_{j = 0}^{n-1} e^{i\frac{2\pi}{n}j} |\psi_{2j+1}(t)|^2,
\label{I}
\end{equation}
and consider its Fourier transform
\begin{equation}
\tilde{\mathcal{I}}(\omega) = \lim\limits_{M \rightarrow \infty} \frac{1}{M}\sum_{t = 0}^{M-1} \mathcal{I}(t) e^{-i \omega t}.
\label{I_til}
\end{equation}
A measure of the time crystallinity is given by
\begin{equation}
Z(t) = e^{-i \frac{2\pi}{n}t} \mathcal{I}(t),
\label{Z}
\end{equation}
whereas the chaoticness of the semiclassical dynamics is quantified by the distance $d^2$ between two slightly different initial states $\vec{\psi}(0) = (1, 0, 0, \dots, 0)^T$ and $\vec{\psi}'(0) = (1-\Delta, \Delta, 0, \dots, 0)^T$ with an arbitrary small $\Delta = 10^{-10}$ 
\begin{equation}
d^2(t) = \sum_{j=1}^{L} (|\psi_j(t)|^2 - |\psi_j'(t)|^2)^2.
\label{d}
\end{equation}
A growth of $d^2(t)$ to a finite value $\sim 1$ corresponds to classical sensitivity to initial conditions and signals quantum thermalization \cite{cosme2014thermalization}.
Finally, we consider the infinite time averages $\langle d^2 \rangle_t = \lim\limits_{M \rightarrow \infty} \frac{1}{M}\sum_{t=0}^{M-1} d^2(t)$ and $\langle Z \rangle_t = \tilde{\mathcal{I}}\left(\frac{2\pi}{n}\right)$.

In the integrable limit $\epsilon = 0$ we have a \gls{DTC}, the phase of $\mathcal{I}$ grows linearly in time, $\tilde{\mathcal{I}}(\omega)$ is peaked at $\omega = \frac{2\pi}{n}$, $\langle Z(t) \rangle_t = 1$ and $\langle d^2 \rangle_t \approx 0$. In the following, we will use these quantities to characterize the dynamical phases of the system.

\begin{figure}[t]
	\begin{center}
		\includegraphics[width=1\linewidth]{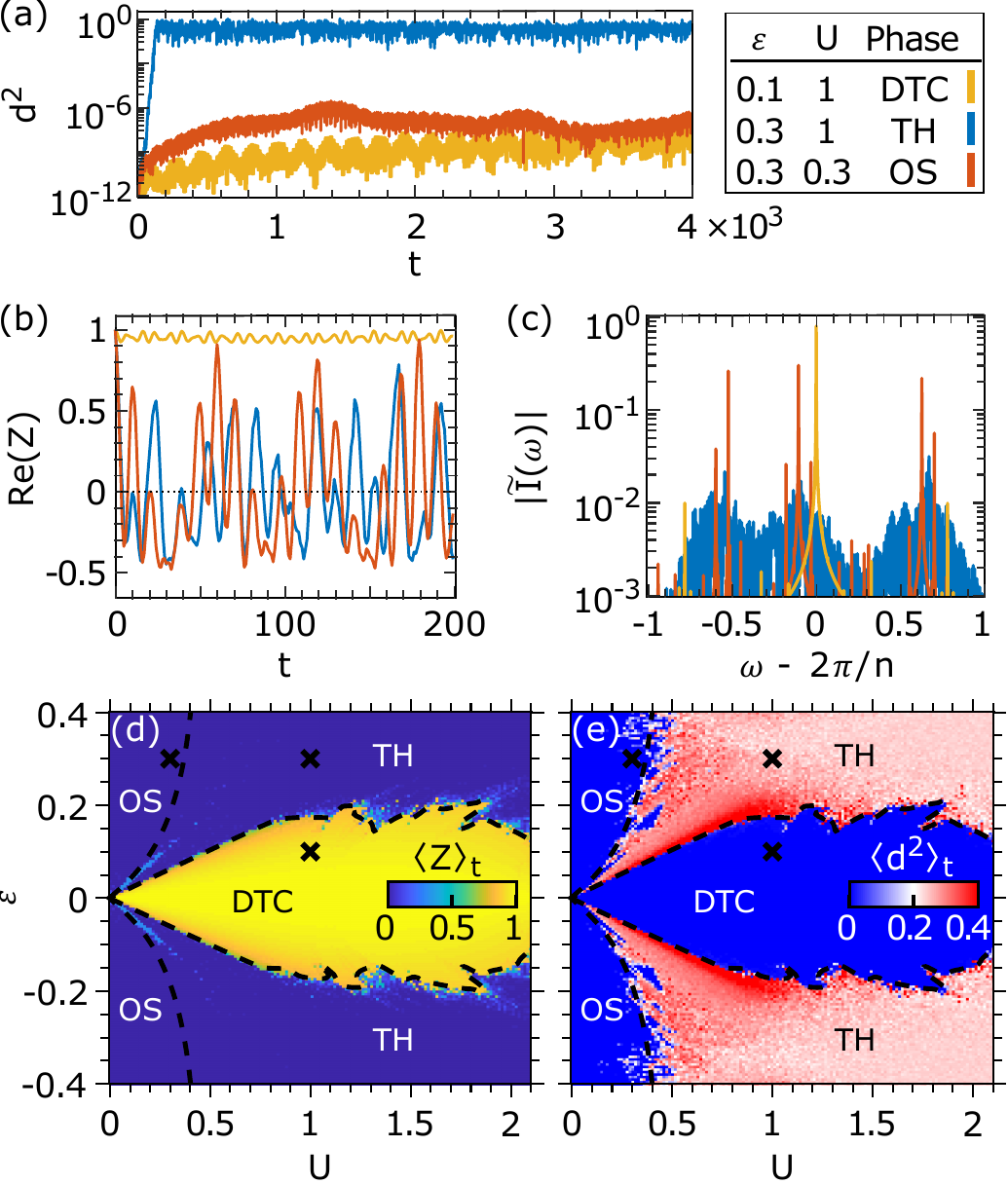}\\
	\end{center}
	\vskip -0.5cm \protect\caption
	{\textbf{Period-$n$ discrete time crystal}, exemplified for $n = L/2 = 3$. For vanishing potential ($h_j = 0$) we characterize the semiclassical dynamics \textit{via} (b) the distance between two slightly different initial states $d^2(t)$ in Eq.~\eqref{d}, (c) the real part of the time crystal order parameter $Z(t)$ in Eq.~\eqref{Z}, and (d) the Fourier transform of the generalized imbalance $\tilde{\mathcal{I}}(\omega)$ in Eq.~\eqref{I_til}.
		On-site interactions ($U$) rigidify the \gls{DTC}, with $Z(t)\approx 1$ and $\tilde{\mathcal{I}}(\omega)$ sharply peaked at $\omega = 2\pi/n$ for small $\epsilon$ (yellow lines).
		For larger $\epsilon$ and small interactions we find an oscillatory dynamical phase (OS), characterized by a few characteristic frequencies at which $\tilde{\mathcal{I}}(\omega)$ is peaked (red lines).
		The thermal phase (TH) has $Z$ oscillating chaotically, $\tilde{\mathcal{I}}(\omega)$ with no dominant peak, and $d^2$ growing to a finite value $\sim 1$ indicating sensitivity to the initial conditions (blue lines).
		The dynamical phases are identified as a function of $\epsilon$ and $U$ \textit{via} $\langle Z \rangle_t$ (e) and $\langle d^2 \rangle_t$ (f), which we compute as time averages over $M=10^4$ Floquet periods. The dashed lines are the same in (e) and (f), the crosses correspond to the parameters considered in (b-d). We observe that interactions rigidify the \gls{DTC}, which expands from $\epsilon = 0$ for increasing $U$.}
	\label{fig: Bosons on ring L=6}
\end{figure}

\textit{Dynamical phases without on-site potential.---}
We start by considering a vanishing on-site potential ($\delta = 0$). In Fig.~\ref{fig: Bosons on ring L=6} we focus on the example $L = 6$, but similar results can be obtained for any $L=2n$ (see SM). We find that three dynamical phases are possible:
(I) period-$n$ \gls{DTC}: for small $\epsilon \approx 0$ the clock-like circulation of the particles is rigidified by the interaction $U \neq 0$ and signaled by $Z \approx 1$. The characteristic frequency of the imbalance $\mathcal{I}$ is locked to $\omega = \frac{2\pi}{n}$, i.e.~the frequency of the peak of $\tilde{\mathcal{I}}$ is robust against perturbations. This is the quintessential property of a \glspl{DTC}: the interactions make the system subharmonic response rigid to mistakes in the driving, that is the range of $\epsilon$ corresponding to the \gls{DTC} grows with $U$. We note that, mathematically, a robust subharmonic response is possible only if the semiclassical dynamical equation is nonlinear, which occurs for non-vanishing interactions $U \neq 0$ and confirms the many-body nature of the \gls{DTC}.
(II) Oscillatory phase: for a large $\epsilon$ but small interaction $U$ the system exhibits non-ergodic oscillations with a few characteristic frequencies signaled by sharp peaks in $\tilde{\mathcal{I}}(\omega)$. In contrast to the case of a \gls{DTC}, these frequencies are not locked, but rather depend on $\epsilon$.
(III) Thermal phase: for large $\epsilon$ and $U$ classical chaos emerges, i.e.~$\langle d^2 \rangle_t \sim 1$. The three phases touch in the tricritical point $\epsilon = U = 0$.

We emphasize that the observation of the \gls{DTC} is not exclusive of the initial condition we considered here but is rather valid for a generic initial Fock state featuring a macroscopic imbalance, thanks to the underlying clock-like particle circulation. Also, we notice that different classes of initial conditions are expected to correspond to \glspl{DTC} with different periods. For instance, an initial state where one site every $2m$ is equally occupied would lead to period $m$-tupling (independently of $L$). In this case, a finite-density occupation could still realize a \gls{DTC} for $L \rightarrow \infty$ \cite{mizuta2018spatial}. Moreover, in general our model does not require any disorder since in a fully connected system (to which our model can be mapped) disorder is not a necessary ingredient to prevent thermalization under Floquet driving \cite{russomanno2017floquet, surace2018floquet}. This is due to an extensive number of integrals of motion as explained in detail in the SM. Finally, we note that the effective full-connectivity of the equivalent clock model ultimately emerges from the symmetry of the multi-particle bosonic wavefunction of the considered physical system of cold atoms, and is in this sense an intrinsic property of the system which cannot be broken by perturbations to the bosonic Hamiltonian.

The realization of period-$n$ DTC in a simple bosonic model in a ring lattice is the first major finding of this work.

\begin{figure*}[t]
	\begin{center}
		\includegraphics[width=\linewidth]{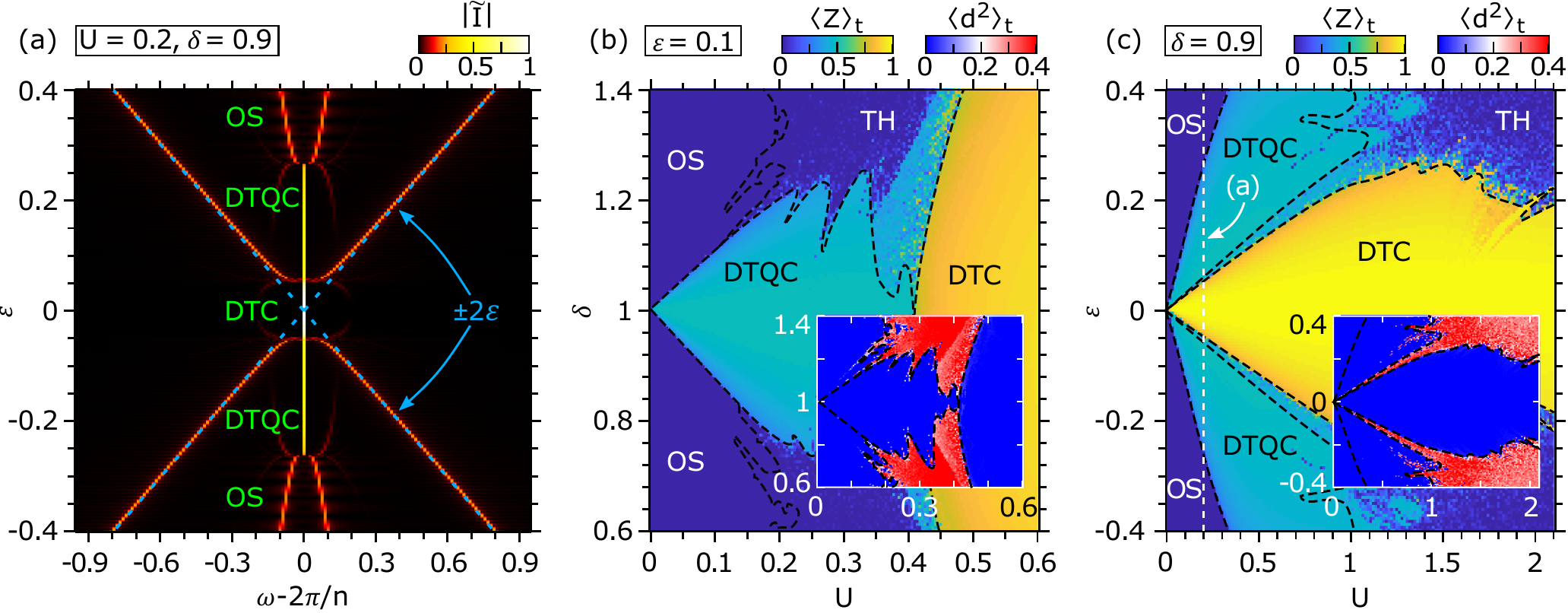}\\
	\end{center}
	\vskip -0.5cm \protect\caption
	{\textbf{Discrete time quasicrystal.} The on-site staircase-like potential leads to a \gls{DTQC}, here exemplified for $L=4$. (a) The dynamical phases are characterized by the Fourier transform of the generalized imbalance $\tilde{\mathcal{I}}(\omega, \epsilon)$ for $\delta = 0.9$, $U = 2$, and $500$ Floquet periods. For a small $|\epsilon|$ we observe the \gls{DTC}, corresponding to a single sharp peak locked at frequency $\frac{2\pi}{n}$. For larger $\epsilon$ we observe the \gls{DTQC}, reflected by a sharp peak locked at frequency $\frac{2\pi}{n}$ and two further peaks at frequencies $\omega_n^{1,2} \approx \pi \pm 2 \epsilon$ (dashed blue lines). (b) The dynamical phases are characterized \textit{via} the quantities $\langle Z \rangle_t = \tilde{\mathcal{I}}(\frac{2\pi}{n})$ and $\langle d^2 \rangle_t$ (inset) computed over $10^4$ Floquet periods and for $\epsilon = 0.1$. The \gls{DTC}, \gls{DTQC}, and oscillatory phases correspond to $\langle Z \rangle \approx 1, 0.5 , 0$, respectively, and thermalization is signaled by a finite $\langle d^2 \rangle_t \sim 1$. Dashed lines serve as a reference and are the same in the main plot and in the inset. The \gls{DTQC} is rigidified by interactions, expanding from $\delta = 1$ for increasing $U$. For sufficiently large $U$, the system enters a \gls{DTC}. (c) We characterize the system as a function of $\epsilon$ and $U$ for $\delta = 0.9$. The minimum interaction required to enter the \gls{DTC} phase increases with $|\epsilon|$, and no \gls{DTC} is possible at all if $|\epsilon|$ is too large.}
	\label{fig: Bosons on ring L=4 QTC}
\end{figure*}

\textit{Tilted lattice and discrete time quasicrystal.---}
Next, we show that tilting the lattice with a staircase-like potential ($\delta \approx 1$) favors a new \gls{DTQC} phase. In presence of such a potential, we gain a clear intuition of the dynamics solving exactly the limit of $U = 0$ and $\delta = 1$ (with $\epsilon$ generally $\neq 0$). In such a limit, the system's characteristic frequencies are linked to the eigenvalues of $F$ and turn out to be (details in the SM) $\frac{2\pi}{n}$, $\omega_n^1$ and $\omega_n^2$. Only the first frequency is locked to a submultiple of the driving frequency, whereas the other two are generally incommensurate with it and are given by
\begin{equation}
\omega_n^{1,2} = \frac{2\pi}{n} \pm \frac{2 \arccos p_n(\epsilon) }{n},
\label{Eq: omega_n^12}
\end{equation}
where $p_n(\epsilon)$ is a trigonometric polynomial in $\epsilon$ with $p_n(0) = 1$,  e.g.~$p_2(\epsilon) = \cos 2\epsilon$ and $\omega_2^{1,2} = \pi \pm 2\epsilon$ for $L = 4$. In the limit $\epsilon = 0$ we find $\omega_n^{1,2} = \frac{2\pi}{n}$, which consistently links back to the period-$n$ \gls{DTC}.
In the thermodynamic limit -- in our case corresponding to a macroscopic number of bosons $N\to \infty$ but finite $L$ -- this regime represents a proper dynamical phase, which we will refer to as a \gls{DTQC} \cite{lifshitz2003quasicrystals, lifshitz2011symmetry}.
The discovery of a \gls{DTQC} and of its characteristic frequencies (Eq.~\eqref{Eq: omega_n^12}) represents the second major finding of this work.
We emphasize that, differently from previous studies \cite{dumitrescu2018logarithmically, peng2018time, zhao2019}, the driving that we consider is perfectly periodic and characterized by a single frequency whereas the system response features a subharmonic frequency $\frac{2\pi}{n}$ \textit{and} two intrinsic frequencies $\omega_n^{1,2}$ that depend on the microscopic parameters of the system.

We show that these results are not an artifact of fine-tuning, but rather underline a proper dynamical phase. A non-zero interaction ($U \neq 0$) indeed rigidifies the \gls{DTQC}. This means that, even if the frequencies $\omega_n^{1,2}$ depend on the model parameters (particularly on $\epsilon$), the characteristic subharmonic frequency $\frac{2\pi}{n}$ does not shift. For $\delta \approx 1$ we observe the new \gls{DTQC} phase, which is characterized by three main sharp peaks in $\tilde{\mathcal{I}}$ (Fig.~\ref{fig: Bosons on ring L=4 QTC}a). One peak is locked to $\frac{2\pi}{n}$ and does not change for $\epsilon$ and $\delta$ within a certain range, whereas the other two are well approximated by the non-interacting prediction of Eq.~\eqref{Eq: omega_n^12}.
We emphasize that, differently from previous works \cite{autti2018observation}, we believe the presence of two incommensurate frequencies to be a necessary but not sufficient ingredient for the definition of a \gls{DTQC} \footnote{A definition relying solely on the presence of two incommensurate frequencies would in fact classify a single, undamped, driven harmonic oscillator as a \gls{DTQC} whenever the driving frequency $\omega_d$ is incommensurate with the oscillator's characteristic frequency $\omega_c$.}. Rather, in analogy with the more established notion of \gls{DTC}, we judge essential the presence of the robust subharmonic peak, which is a genuine many-body feature.
Beyond the \gls{DTQC}, we also still observe the \gls{DTC}, oscillatory, and chaotic phases, which are reflected in $\tilde{\mathcal{I}}(\omega)$ featuring a single peak locked to $\frac{2\pi}{n}$, few sharp peaks and no sharp peaks at all, respectively. In Fig.~\ref{fig: Bosons on ring L=4 QTC}b we sketch the various phases as a function of $\delta$ and $U$  for $\epsilon = 0.1$ by looking at $\langle Z \rangle_t = \tilde{\mathcal{I}}(\frac{2\pi}{n})$ and $\langle d^2 \rangle_t$. The interaction $U$ rigidifies the \gls{DTQC}, which expands from $\delta = 1$ for increasing $U$ and is signaled by $\langle Z \rangle \approx 0.5$. For increasing interactions the system enters a thermal and/or a \gls{DTC} phase. In Fig.~\ref{fig: Bosons on ring L=4 QTC}c we characterize the system as a function of $\epsilon$ and $U$ for $\delta = 0.9$. We see that the interaction necessary to enter the \gls{DTC} grows with $\epsilon$. Eventually, if $\epsilon$ is too large, no \gls{DTC} is possible.

\textit{Experimental implementations.---}
An appealing feature of our proposal lies in its ultimate simplicity making it very natural for experimental implementation. Indeed, the Hamiltonian \eqref{eq: H ternary} only requires a number of basic ingredients such as nearest-neighbor hopping, local interaction and on-site potential, which are all readily available for ring-shaped optical lattices \cite{franke2007optical, houston2008reproducible, vyas2013polarization, amico2014superfluid}. The main difficulty is the discrete time dependent switch of tunnelling between alternating bonds. For small systems this is easily achievable if alternating bonds point in different real space directions. Alternatively, the even and odd site labels could be imprinted onto two different spin states. This would enable the discrete switch also on larger ring lattices. More concretely,  the dynamics of the first and of the second third of the Floquet Hamiltonian would in this case correspond to a transverse field-induced spin-flip and to a intra-well barrier lowering, respectively. Furthermore, we have also checked that our results are robust against small undesired hoppings at odds with the alternating Floquet protocol.

\textit{Conclusions.---}
In conclusion, we have studied the non-equilibrium properties of interacting bosons in a ring lattice. Considering a Floquet driving which alternates hopping on even and odd nearest-neighbor links, we induced a clock-like circulation of the particles whose direction depends on the parity of sites. Introducing a suitable set of dynamical order parameters and solving the exact dynamics in two integrable limits and a Gross-Pitaevskii equation across the whole parameter space, we identified several dynamical phases: a period-$n$ \gls{DTC}, an oscillatory phase, a thermal phase, and a new \gls{DTQC}. The \gls{DTQC} phase is characterized by a frequency which is locked to a submultiple of the Floquet frequency and by a second frequency which is incommensurate with it. Our work demonstrates a wide spectrum of non-equilibrium, non-trivial dynamical phases of bosons, which represent a unique opportunity for experimental investigation.

As natural for bosons in a finite-size lattice, the proposed system can effectively be mapped onto a fully-connected model. A natural question for future investigation regards then the fate of the various dynamical phases in physical (rather than effective) systems of clocks where such a full-connectivity is broken, e.g.~in presence of long-range power-law interaction. Moreover, our work adds a totally new perspective to the multiple connotations of the still developing concept of time quasi-crystallinity, which certainly deserves further study. For instance, an intriguing question concerns the possibility for a system driven with two incommensurate frequencies $\omega_1$ and $\omega_2$ to respond at subharmonic frequencies $\omega_1/n_1$ and $\omega_2/n_2$.

\begin{acknowledgments}
	It is a pleasure to thank N.~R.~Cooper, J.~P.~Garrahan, M.~Landini and A.~Lazarides for useful discussions.
	A.~P.~acknowledges support from the Royal Society. A.~N.~holds a University Research Fellowship from the Royal Society and acknowledges additional support from the Winton Programme for the Physics of Sustainability.
\end{acknowledgments}

\bibliographystyle{apsrev4-1}
\bibliography{Bosons_on_ring_biblio}


\newpage

\setcounter{equation}{0}
\setcounter{figure}{0}
\setcounter{page}{1}
\makeatletter 
\renewcommand{\thefigure}{S\arabic{figure}}

\onecolumngrid

\begin{center}
	{\fontsize{12}{12}\selectfont
		\textbf{Supplemental Material for "Period-$n$ discrete time crystals and quasicrystals with ultracold bosons"\\[5mm]}}
	{\normalsize Andrea Pizzi, Johannes Knolle and Andreas Nunnenkamp \\[1mm]}
\end{center}
\normalsize

\section{I) Dynamical equations and solvable limits}

We derive explicitly the quantum dynamical equation and solve it in the limit cases $\epsilon = 0$ and $U = 0$. The Heisenberg dynamical equations for the annihilation operator $a_{j}$ at site $j = 1, 2, \dots, L = 2n$ is obtained with a straightforward computation of the commutators of the Hamiltonian (1) with $a_{j}$ and reads ($\hbar = 1$)
\begin{equation}
\frac{d a_j}{dt} = 
\begin{cases}
i 3 J_1 a_{j \pm 1} \quad & 0 < t \mymod{T} < \frac{1}{3} \\
i 3 J_2 a_{j \mp 1} \quad & \frac{1}{3} < t \mymod{T} < \frac{2}{3} \\
-i 3 \left[\frac{U}{N}n_j + h_j\right] \quad & \frac{2}{3} < t \mymod{T} < 1, \\
\end{cases}
\label{eq: Heisenberg}
\end{equation}
where the upper and the lower signs are for odd and even $j$, respectively. In the following we implicitly assume a stroboscopic time $t = 0,1,2,\dots$, unless differently specified. Recalling $J_1 = J_2 = \frac{\pi}{2} + \epsilon$, we integrate the dynamics over the first two fractions of the Floquet driving to obtain
\begin{align}
\begin{pmatrix}
a_{2j} \\
a_{2j-1}
\end{pmatrix} \left(t + 1/3 \right) &=
-\begin{pmatrix}
\sin \epsilon & i\cos \epsilon \\
i\cos \epsilon & \sin \epsilon
\end{pmatrix}
\begin{pmatrix}
a_{2j} \\
a_{2j-1}
\end{pmatrix} \left(t\right),
\\
\begin{pmatrix}
a_{2j} \\
a_{2j+1}
\end{pmatrix} \left(t + 2/3\right) &=
-\begin{pmatrix}
\sin \epsilon & i\cos \epsilon \\
i\cos \epsilon & \sin \epsilon
\end{pmatrix}
\begin{pmatrix}
a_{2j} \\
a_{2j+1}
\end{pmatrix} \left(t + 1/3\right),
\end{align}
whereas the third fraction of the Floquet driving can be integrated only in the two limit cases $\epsilon = 0$ and $U = 0$.

\subsection{Limit $\epsilon = 0$}
For a fine-tuned hopping strength with $\epsilon = 0$, we get
\begin{align}
\begin{pmatrix}
a_{2j} \\
a_{2j-1}
\end{pmatrix} \left(t + 1/3\right) &=
- i
\begin{pmatrix}
a_{2j-1} \\
a_{2j}
\end{pmatrix} \left(t \right),
\\
\begin{pmatrix}
a_{2j} \\
a_{2j+1}
\end{pmatrix} \left(t + 2/3\right) &=
-i
\begin{pmatrix}
a_{2j+1} \\
a_{2j}
\end{pmatrix} \left(t + 1/3\right)
=
-
\begin{pmatrix}
a_{2j+2} \\
a_{2j-1}
\end{pmatrix} \left(t \right),
\end{align}
and thus
\begin{equation}
\begin{aligned}
n_{2j}(t+2/3) &= n_{2j+2}(t), \\
n_{2j+1}(t+2/3) &= n_{2j-1}(t). \\
\end{aligned}
\end{equation}
Moreover, since $[H, n_j] = 0$ for $2<t<3$, we finally get
\begin{equation}
n_{j}(t) = n_{j \mp 2t}(0),
\end{equation}
where again the upper and lower sign refer to odd and even $j$, respectively. We thus obtain that $n_j(t = n) = n_j(t = 0)$, that is the solvable limit $\epsilon = 0$ is at the core of a period-$n$ \gls{DTC}.

\subsection{Limit $U = 0$}
In the non-interacting limit ($U=0$), also the dynamics associated to the third fraction of the Floquet period becomes linear, and can be trivially solved by $a_j(t+1) = e^{-ih_j} a_j(t+2/3)$. The evolution of the bosonic operators $\vec{a} = (a_1, a_2, \dots, a_{2n})^T$ over one period can be compactly written as $\vec{a}(t+1) = F \vec{a}(t)$ with
\begin{equation}
\begin{aligned}
F & = \Pi_h \left[-\sin (\epsilon) \mathbb{1}_{2n} + i \cos (\epsilon) K_2\right]	\left[-\sin (\epsilon) \mathbb{1}_{2n} + i \cos (\epsilon) K_1\right] \\
&=\Pi_h \left[\sin(\epsilon)^2 \mathbb{1}_{2n} - \cos(\epsilon)^2 K_2 K_1 -i\sin(\epsilon)\cos(\epsilon) (K_1 + K_2)\right],
\end{aligned}
\end{equation}
where $\mathbb{1}_{2n}$, $K_1$, $K_2$ and $\Pi_h$ are $2n \times 2n$-dimensional matrices. In particular, $\Pi_h$ is diagonal and with entries $(\Pi_h)_{j,j} = e^{-ih_j}$, $\mathbb{1}_{2n}$ is the identity and $K_1$ and $K_2$ are $2n \times 2n$-dimensional involution matrices swapping sites $(1,2), (3,4), \dots, (L-1,L)$ and sites $(2,3), (4,5), \dots, (L, 1)$, respectively
\begin{equation}
K_1=
\begin{pmatrix}
0 & 1 &   &   &   &   &   \\
1 & 0 &   &   &   &   &   \\
&   & 0 & 1 &   &   &   \\
&   & 1 & 0 &   &   &   \\
&   &   &   & \ddots &   &  \\
&   &   &   &   & 0 & 1 \\
&   &   &   &   & 1 & 0 \\	  
\end{pmatrix},
\quad
K_2=
\begin{pmatrix}
0 &   &   &   &   &   & 1 \\
& 0 & 1 &   &   &   &   \\
& 1 & 0 &   &   &   &   \\
&   &   & 0 & 1 &   &   \\
&   &   & 1 & 0 &   &   \\
&   &   &   &   & \ddots &  \\
1 &   &   &   &   &   & 0 \\	  
\end{pmatrix},
\end{equation}
It is easy to check that $K_2 K_1 = \tau_e^\dagger + \tau_o$ where $\tau_e$ and $\tau_o$ are defined by	
\begin{align}
&(\tau_e)_{2i, 2j+1} = (\tau_e)_{2i+1, 2j} = 0, \\
&(\tau_o)_{2i, 2j+1} = (\tau_o)_{2i+1, 2j} = 0, \\
&(\tau_e)_{2i, 2j} = (\tau_o)_{2i+1, 2j+1} = \tau,
\end{align}	
with $\tau$ the following $n \times n$-dimensional matrix

\begin{equation}
\tau = 
\begin{pmatrix}
0 &   & & & 1\\
1 & 0 & & & &\\
&\ddots &\ddots& & &\\
&   &     1 &0& &\\
&   &       &1& 0\\
\end{pmatrix}.
\end{equation}

The matrix $K_2 K_1$ thus generates a clockwise (counterclockwise) circulation of the particles within the odd (even) sites.

In general, the system dynamics is characterized by $2n$ frequencies which are linked to the eigenvalues of $F$. Considering the staircase-like potential $h_{2j} = h_{2j-1} = \frac{2\pi}{n}j \delta$, these eigenvalues take a particular form in the limit $\delta = 1$, which we are about to show and which is at the core of the \gls{DTQC} phase.

Let us say $\omega_j = e^{i\frac{2\pi}{n}j}$, $\alpha_j = \omega_j \sin(\epsilon)^2$, $\beta_j = - \omega_j \cos(\epsilon)^2$, $\gamma_j = - i \omega_j \cos(\epsilon) \sin(\epsilon)$ and write $F$ as
\begin{equation}
F = 
\begin{pmatrix}
\alpha_1 & \gamma_1                    & & & & & & & \beta_1 & \gamma_1\\
\gamma_1 & \alpha_1 & \gamma_1 & \beta_1   & & & & & & \\
\beta_2  & \gamma_ 2& \alpha_2 & \gamma_2  & & & & & & \\
& & \gamma_2 & \alpha_2 & \gamma_2 & \beta_2   & & & & \\
& & \beta_3 & \gamma_3 & \alpha_3 & \gamma_3   & & & & \\
& & & & & \ddots                               & & & & \\
& & & & \beta_{n-1} & \gamma_{n-1} & \alpha_{n-1} & \gamma_{n-1} \\
& & & & & & \gamma_{n-1} & \alpha_{n-1} & \gamma_{n-1} & \beta_{n-1} \\
& & & & & & \beta_n & \gamma_n & \alpha_n & \gamma_n \\
\gamma_n & \beta_n & & & & & & & \gamma_n & \alpha_n \\
\end{pmatrix}.
\end{equation}

\begin{figure}[t]
	\begin{center}
		\includegraphics[width=\linewidth]{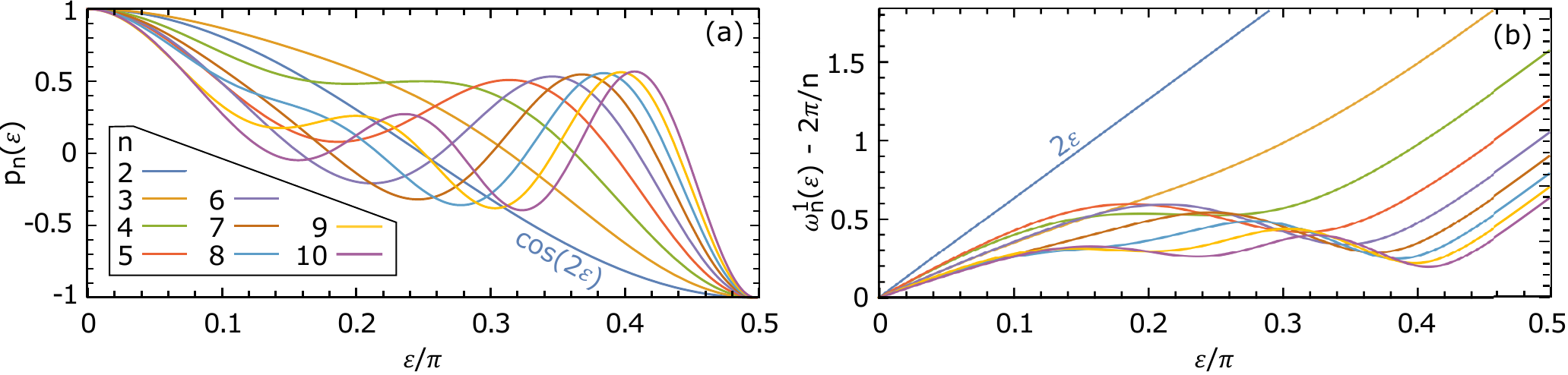}\\
	\end{center}
	\vskip -0.5cm \protect\caption
	{\textbf{Characteristic frequency of a \gls{DTQC}}. Tilting the lattice with a staircase-like potential $h_{2j-1} = h_{2j} = \frac{2\pi}{n}j$, the eigenvalues of the evolution matrix $F$ are described in terms of a trigonometric polynomial $p_n(\epsilon)$, which is thus connected to the frequencies $\omega_n^{1,2}$ characterizing the \gls{DTQC}. (a) We plot $p_n(\epsilon)$ for $n = 2, 3, \dots, 10$. Notice that $p_n(0) = 1$ and that for a 4-site lattice we get $p_2(\epsilon) = \cos(2\epsilon)$. In particular, for $L = 4$ sites, we find $p_2(\epsilon) = \cos(2 \epsilon)$. (b) We plot the corresponding frequency $\omega_n^{1}(\epsilon) - \frac{2\pi}{n} = \frac{2\pi}{n} - \omega_n^{2}(\epsilon)= \frac{2}{n} \arccos(p_n(\epsilon))$. For $\epsilon = 0$ we get $\omega_n^{1,2} = \frac{2\pi}{n}$, recovering the \gls{DTC} as expected.}
	\label{fig: trigo poly & characteristic frequencies}
\end{figure}

The eigenvalue problem reads\
\begin{equation}
F \vec{y} = \lambda \vec{y},
\label{eq:eig problem.1}
\end{equation}
where $\vec{y} = (y_1^o, y_1^e, y_2^o, y_2^e, \dots, y_n^o, y_n^e)^T$ is a $2n$-dimensional column vector. We rewrite the eigenvalue problem \eqref{eq:eig problem.1} component by component as
\begin{equation}
\begin{cases}
(F \vec{y})_{2j}   &= \omega_j \big( \gamma y_j^o + \alpha y_j^e + \gamma y_{j+1}^o + \beta y_{j+1}^e \big) = \lambda y_j^e \\
(F \vec{y})_{2j-1} &= \omega_j \big( \gamma y_j^e + \alpha y_j^o + \gamma y_{j-1}^e + \beta y_{j-1}^o \big) = \lambda y_j^o. \\
\end{cases}
\label{eq:eig problem.2}
\end{equation}
Dividing both members of Eqs.~\eqref{eq:eig problem.2} by $\omega_j$ and introducing the Fourier transform $\tilde{y}_k^{e/o} = \sum_{j=1}^{n} \omega_j y_{j}^{e/o}$ (with $k = 1, 2, \dots, n$) we get
\begin{equation}
\begin{cases}
\gamma \tilde{y}_k^o + \alpha \tilde{y}_k^e + \frac{\gamma}{\omega_k} \tilde{y}_k^o + \frac{\beta}{\omega_k} \tilde{y}_k^e = \lambda \tilde{y}_{k-1}^e \\
\gamma \tilde{y}_k^e + \alpha \tilde{y}_k^o + \gamma \omega_k \tilde{y}_k^e + \beta \omega_k \tilde{y}_k^o = \lambda \tilde{y}_{k-1}^o, \\
\end{cases}
\label{eq:eig problem.3}
\end{equation}
that we rewrite in a compact form as
\begin{equation}
\lambda 
\begin{pmatrix}
\tilde{y}_k^e \\
\tilde{y}_k^e
\end{pmatrix}
= M_k
\begin{pmatrix}
\tilde{y}_{k-1}^e \\
\tilde{y}_{k-1}^e
\end{pmatrix},
\label{eq:eig problem.4}
\end{equation}
where $M_k$ reads
\begin{equation}
M_k =
\begin{pmatrix}
\sin(\epsilon)^2 - \frac{\cos(\epsilon)^2}{\omega_k} & -i \sin(\epsilon) \cos(\epsilon) (1+ \frac{1}{\omega_k}) \\
-i \sin(\epsilon) \cos(\epsilon) (1+ \omega_k) & \sin(\epsilon)^2 - \cos(\epsilon)^2 \omega_k
\end{pmatrix}.
\end{equation}
It is easy to show that $\det(M_k) = 1$ for every $k$, so that $M_k$ admits an inverse $M_k^{-1}$. Inverting Eq.~\eqref{eq:eig problem.4}, iterating it and exploiting periodicity in momentum space ($\tilde{y}_k^{e/o} = \tilde{y}_{k \pm n}^{e/o}$) we get
\begin{equation}
\begin{pmatrix}
\tilde{y}_{k}^e \\
\tilde{y}_{k}^e
\end{pmatrix}
=
\lambda^n M_{k+1}^{-1} M_{k+2}^{-1} M_{k+3}^{-1} \dots M_{k+n-1}^{-1} M_{k}^{-1}
\begin{pmatrix}
\tilde{y}_{k}^e \\
\tilde{y}_{k}^e
\end{pmatrix},
\end{equation}
which is itself a $2 \times 2$-dimensional eigenvalue problem for the eigenvalue $\lambda^n$. The matrix $M_n M_{n-1} \dots M_1$ has determinant $1$, and its eigenvalues can therefore be written as $- e^{\pm i \acos(-\frac{\Tr}{2})}$ where $\Tr$ is its trace. For $\delta = 1$, we thus finally get the eigenvalues of $F$ to be
\begin{equation}
\lambda_j^{\pm} = \exp \left[i \frac{\pi + 2 \pi j \pm \arccos p_n(\epsilon)}{n}\right],
\label{lam}
\end{equation}
where
\begin{equation}
p_n(\epsilon) = - \frac{1}{2} \sum_{j_1, j_2, \dots, j_n = 1}^{2} (M_n)_{j_1, j_2} (M_{n-1})_{j_2, j_3} \dots (M_1)_{j_n, j_1}
\end{equation}
is in general a trigonometric polynomial in $\epsilon$, of order $2n$ and such that $p_n(0) = 1$. For instance, for $L = 2n = 4$ we obtain
\begin{equation}
M_1 =
\begin{pmatrix}
1 & 0 \\
0 & 1
\end{pmatrix},
\quad \quad
M_2 =
\begin{pmatrix}
\sin(\epsilon)^2 - \cos(\epsilon)^2 & -i 2\sin(\epsilon) \cos(\epsilon) \\
-i 2\sin(\epsilon) \cos(\epsilon) & \sin(\epsilon)^2 - \cos(\epsilon)^2
\end{pmatrix},
\end{equation}
so that $\Tr = 2\sin(\epsilon)^2 - 2\cos(\epsilon)^2$, i.e.~$p_2(\epsilon) = \cos(2\epsilon)$. We plot $p_n(\epsilon)$ for $n = 2, 3, \dots, 10$ in Fig.~\ref{fig: trigo poly & characteristic frequencies}a, and report here its explicit expression for $n = 2,3,\dots, 6$
\begin{align}
p_2(\epsilon) = &\cos(2\epsilon),\\
p_3(\epsilon) = &\frac{1}{16} \left(3 + 15 \cos(2\epsilon) - 3 \cos(4\epsilon) + \cos(6\epsilon)\right),\\
p_4(\epsilon) = &\frac{1}{2} \left(1 + 3 \cos(2\epsilon) - \cos(4\epsilon) + \cos(6\epsilon)\right), \\
p_5(\epsilon) = &\frac{1}{256} \bigg( 95-20\sqrt{5} - 10(\sqrt{5}-16) \cos(2\epsilon) + 20(\sqrt{5}-4) \cos(4\epsilon) + \dots \\
& + 5(2\sqrt{5}+19) \cos(6\epsilon) - 15 \cos(8\epsilon) + \cos(10\epsilon)\bigg), \\
P_6(\epsilon) = &\frac{1}{256} \bigg( 39 + 114 \cos(2\epsilon) + 12 \cos(4\epsilon) + 133 \cos(6\epsilon) - 51 \cos(8\epsilon) + 9 \cos(10 \epsilon) \bigg).
\end{align}
As explained in the main text, the eigenvalues \eqref{lam} are at the core of a \gls{DTQC} dynamical phase. The characteristic frequencies $\omega_n^{1,2}$ are plotted in Fig.~\ref{fig: trigo poly & characteristic frequencies}b. In the limit $\epsilon \rightarrow 0$ we get $\omega_n^{1,2} \rightarrow \frac{2\pi}{n}$, consistently recovering the \gls{DTC}.

\section{II) Map to a fully connected clock model}
Our bosonic model is equivalent to a model of fully connected (i.e.~infinite-range interacting) clock variables of two counter-rotating species, as we show here in the limit $\epsilon = 0$. For simplicity we adopt the inverse route: we present the clock model and map it back to the bosonic one, following in spirit the mapping that Ref.~\cite{surace2018floquet} carried out for clocks of a single species.

Consider a system of $N$ clock variables with $n$-hands, which generalizes the spin case corresponding to $n=2$. To each clock we further associate a binary label $s = e, o$ indicating the species of the clock. We will refer to clocks of species $e,o$ as to \textit{even} and \textit{odd} clocks, respectively. A natural basis of the Hilbert space is the one of states $\ket{\{s_i, j_i\}} = \bigotimes_{i=1}^{N} \ket{s_i, j_i}_i$ with the $i$-th clock being of species $s_i$ and in the $j_i$-th hand.
We introduce the local operators $\sigma_{s,i}$, $\eta_{s,i}$ and $\tau_{s,i}$ acting on the $i$-th clock and defined as
\begin{align}
\sigma_{s,i} &= \sum_{j = 1}^{n} \ket{s, j}_i \omega^{j} \bra{s, j}_i, \\
\eta_{s,i}   &= \sum_{j = 1}^{n} \ket{s, j}_i h_{s, j} \bra{s, j}_i, \\
\tau_{s,i}   &= \sum_{j = 1}^{n} \ket{s, j + 1}_i \bra{s, j}_i,
\end{align}
where $\bar{s} = e,o$ for $s = o,e$, respectively, $\omega = e^{i\frac{2\pi}{n}}$, and where $\{h_{s,j}\}$ are so far unspecified numbers. The operator $\tau_{s,i}$ acts on the $i$-th clock moving its hand one step forward, if of species $s$, or annihilating it, is of species $\bar{s}$. Consider a Floquet operator $U_F = e^{-iH} U_K$, where $H$ is the Hamiltonian
\begin{equation}
H = \frac{U}{NL} \sum_{m=0}^{n-1} \sum_{i_1,i_2=1}^{N} \sum_{s=e,o} (\sigma_{s,i_1}^\dagger \sigma_{s,i_2})^m + \sum_{i=1}^{N} \sum_{s = e,o} \eta_{s,i},
\label{eq: clocks H}
\end{equation}
and $U_K$ is a local kicking operator of the form
\begin{equation}
U_K = \prod_{i=1}^{N}(\tau_{e,i}^\dagger + \tau_{o,i}).
\label{eq: kicking operator}
\end{equation}

The first term of the Hamiltonian \eqref{eq: clocks H} plays the role of interaction among clocks of the same species and couples all sites, making the system fully connected. In the chosen basis, $H$ is diagonal whereas the operator $U_K$ acts moving one step forward (backward) the hands of the odd (even) clocks. After $n$ Floquet periods, the state of the system will therefore come back to the initial condition. This mechanism is at the basis of a period-$n$ \gls{DTC}.

Being fully connected, the system can be described just counting the number of clocks of a certain species and with hands pointing in a certain direction. That is, it is possible to describe the system in terms of bosonic operators $b_j, b_j^\dagger$ with $j = 1, 2, \dots, 2n$ and fulfilling the standard bosonic commutation relations $[b_j, b_{j'}^\dagger] = \delta_{j, j'}$ and $[b_j, b_{j'}] = 0$ (for further details we refer to Appendix D of \cite{surace2018floquet}). We say
\begin{equation}
\begin{aligned}
\ket{n_1, n_2, \dots, n_{2n}} = \frac{1}{\sqrt{N! \prod_{j=1}^{2n} n_j!}} P
\bigg(
&\underbrace{\ket{o,1}_1 \otimes \ket{o,1}_2 \otimes \dots \otimes \ket{o,1}_{n_1}}_{n_1 \text{times}}
\otimes
\underbrace{\ket{e,1}_{n_1+1} \otimes \ket{e,1}_{n_1+2} \otimes \dots \otimes \ket{e,1}_{n_1+n_2}}_{n_2 \text{times}} \\
&\otimes\dots\otimes
\underbrace{\ket{e,n}_{N-n_{2n}+1} \otimes \ket{e,n}_{N-n_{2n}+2} \dots \otimes \ket{e,j}_N}_{n_{2n} \text{times}}
\bigg)
\end{aligned}
\end{equation}
where $P = \sum_{i_1, i_2, \dots, i_N} \Pi_{i_1, i_2, \dots, i_N}$ is the symmetrization operator, with sum running over the possible permutations of $1, 2, \dots, N$ and with $\Pi_{i_1, i_2, \dots, i_N}$ being the corresponding permutation operator. The bosonic operators are such that
\begin{equation}
\ket{n_1, n_2, \dots, n_{2n}} = \frac{1}{\sqrt{\prod_{j=1}^{2n} n_j!}} (b_1^\dagger)^{n_1} (b_2^\dagger)^{n_2} \dots (b_{2n}^\dagger)^{n_{2n}} \ket{\text{vac}}
\end{equation}

Since the Hamiltonian \eqref{eq: clocks H} is invariant under site permutations, it is easy to show that
\begin{equation}
H = \frac{U}{NL} \sum_{m=0}^{n-1} \sum_{j_1, j_2 = 1}^{2n} n_{j_1} n_{j_2} \omega^{(j_2-j_1) m} + \sum_{j=1}^{n} (h_{o, j} n_{2j-1} + h_{e, j} n_{2j}),
\label{eq: clock H map.1}
\end{equation}
where $n_j = b_j^\dagger b_j$ is the number operator for the $j$-th bosonic mode. From Eq.~\eqref{eq: clock H map.1}, performing the sum over $m$, removing a constant term $U$ and saying $h_{2j-1} = h_{o, j}$ and $h_{2j} = h_{e, j}$, we finally obtain
\begin{equation}
H = \sum_{j=1}^{2n} \left[ \frac{U}{N} n_j (n_j-1) + h_j n_j \right].
\end{equation}
As well, we find that the kickting operator acts as
\begin{equation}
U_K \ket{n_1, n_2, n_3, n_4, n_5, \dots, n_{2n}} = \ket{n_{2n-1}, n_4, n_1, n_6, n_3, \dots, n_{2n}}.
\end{equation}

Summing up, we have therefore shown that the fully-connected clock model can be mapped into the model of bosons in a ring with $2n$ sites. The effect of the kicking operator is to rotate the bosons in the odd (even) sites of two steps in the clockwise (countercockwise) direction, whereas the clock Hamiltonian \eqref{eq: clocks H} maps into a local potential and two-body interaction for bosons. In the main, the kicking operator is then realized thanks to the sequential action of first and the second fractions of the ternary Floquet Hamiltonian.

The effective full-connectivity of the bosonic model has an important implication: the system is invariant under the permutation of clocks $(1,2)$, $(2,3)$, $\dots, (N-1, N)$, which corresponds to an \textit{extensive} number $\lfloor \frac{N}{2} \rfloor$ of integrals of motion. Thanks to the existence of these integrals of motion, \textit{a priori} the system does not necessitate any disorder to escape the fate of thermalization, making non-trivial dynamical phases such as \glspl{DTC} possible.

\section{III) On the validity of mean field}
Throughout the main text, numerical results are obtained solving the \gls{GPE}, which is often referred to as the \gls{MF} limit of the Heisenberg equation. Here we discuss the regimes in which this is legitimate, and how to interpret the results when it is not.

In the considered thermodynamic limit ($N \rightarrow \infty$ for a fixed $L$), the correct and well-established phase-space framework to deal with non-equilibrium dynamics is the \gls{TWA} \cite{polkovnikov2003quantum}. This computes expectation values as averages over the classical \gls{GPE} trajectories obtained for an ensemble of classical initial conditions, whose stochastic distribution is chosen consistently with the actual quantum initial condition. When the system is initialized in $\ket{\psi(0)} = \ket{N, 0, \dots, 0}$, the initial condition for the \gls{GPE} reads $\psi_j(0) = \delta_{j, 1} e^{i\theta_j}$ and, thanks to the gauge symmetry, can actually be considered without loss of generality to be $\psi_j(0) = \delta_{j, 1}$. Since the initial condition no longer contains any degree of freedom, in this case the \gls{TWA} is equivalent to a \gls{SSGPE}. This observation sounds very promising from a computational point of view, because it allows avoiding running the \gls{GPE} multiple times, but it comes with a drawback: the \gls{SSGPE} is not guaranteed to be accurate since the initial condition features non-macroscopically occupied sites. In particular, we expect the \gls{SSGPE} to be inaccurate when chaotic. In the framework of \gls{TWA}, chaotic semiclassical equations are in fact expected to underline thermalization \cite{cosme2014thermalization}, whereas the \gls{SSGPE} displays in general persistent (in fact, chaotic) oscillations.

\subsection{Exact diagonalization}
\begin{figure}[t]
	\begin{center}
		\includegraphics[width=	\linewidth]{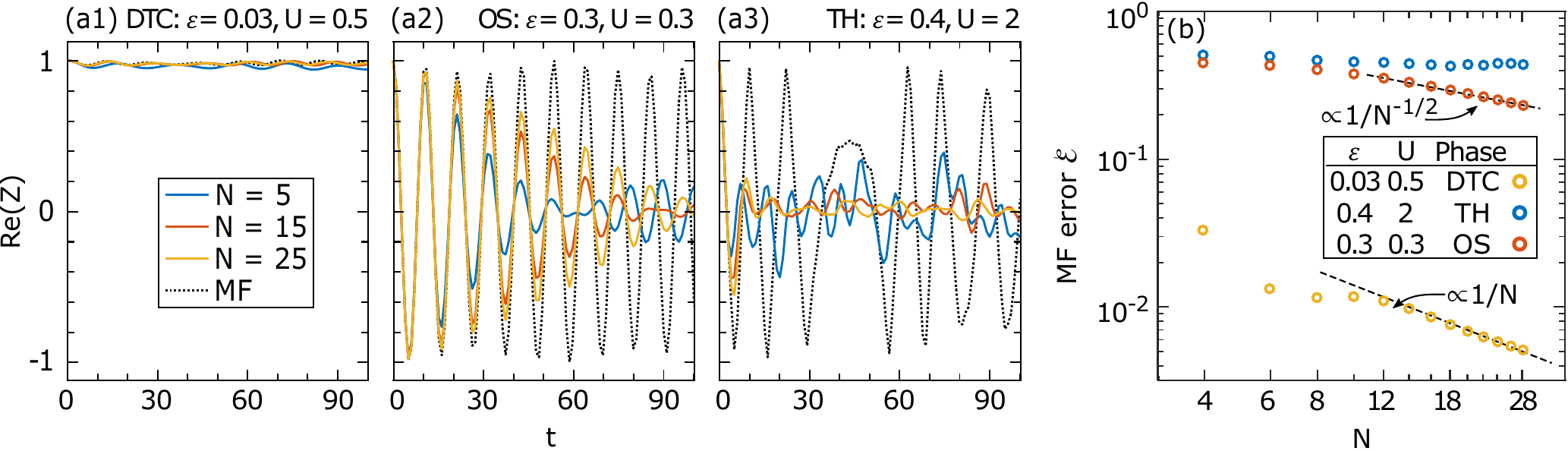}\\
	\end{center}
	\vskip -0.5cm \protect\caption
	{\textbf{Exact diagonalization results}. We present results obtained within \gls{ED} for $L = 4$ and in absence of on-site potential, showing evidence for the consistency of \gls{MF} in the thermodynamic limit. We plot the real part of the time crystallinity order parameter $Z$ (a1-a3), at stroboscopic times, for $N=5, 15, 25$, comparing it with the \gls{MF} result. We observe different trends corresponding to the different dynamical phases. In the \gls{DTC} (a1) the agreement between \gls{ED} and \gls{MF} improves with for growing $N$ and is rather good already for small $N$; in the oscillatory phase (a2) the \gls{MF} accuracy deteriorates in time, yet improves for increasing $N$: in the thermodynamic limit we expect the \gls{MF} results to be exact at all times; in the thermal phase (a3) the \gls{MF} is inaccurate irrespective of $N$, since thermalization quickly leads to $Z_{ED} \approx 0$ whereas $Z_{MF}$ rather fluctuates chaotically. (b) We plot the measure of the error $\mathcal{E}$ of \gls{MF}, see Eq.~\eqref{E}. For the \gls{DTC} and the oscillatory phases, we find that $\mathcal{E}$ decays as a power of $N$ (notice the logarithmic axes and the reference dashed lines $\propto 1/N, 1/N^{1/2}$), whereas in the thermal phase $\mathcal{E}$ does not decay with $N$.}
	\label{fig: ED vs MF}
\end{figure}

To support the previous claims we compare the \gls{SSGPE} with \gls{ED} computations for $L = 4$ sites, finite $N \le 28$ and vanishing on-site potential ($\delta = 0$). For various values of $\epsilon$ and $U$ corresponding to the \gls{DTC}, oscillatory and thermal phases, in Fig.~\ref{fig: ED vs MF}a1-a3 we plot $Z(t)$. The \gls{DTC} is characterized by $Z \approx 1$ already for a relatively small $N$, whereas in the oscillatory phase the accuracy of the \gls{MF} solution is worse and deteriorates in time. Nevertheless, in both cases, for increasing $N$ the accuracy of \gls{MF} improves, and in the thermodynamic limit $N \rightarrow \infty$ (for a fixed, finite $L$) we expect \gls{MF} to be exact up to arbitrarily large times. Conversely, in the thermal phase the \gls{MF} solution is inaccurate irrespective of $N$. In this phase, $Z_{MF}$ fluctuates chaotically whereas $Z_{ED}$ saturates to a steady value, i.e.~thermalizes, as expected \cite{cosme2014thermalization}.

To better interpret the results, we check the finite-size scaling considering the parameter
\begin{equation}
\mathcal{E}(N) = \frac{1}{100} \sum_{t = 0}^{99} |Z_{ED}(t; N) - Z_{MF}(t)|,
\label{E}
\end{equation}
which serves as a measure of the \gls{MF} error with respect to \gls{ED} for a given $N$. In Fig.~\ref{fig: ED vs MF}b we show that $\mathcal{E}$ decays as a power of $N$ in the \gls{DTC} and oscillatory phases, whereas it does not decay in the thermal phase. This supports the idea for which, in the non-chaotic regime and in the thermodynamic limit, the \gls{SSGPE} is exact. On the other hand, a failure of the \gls{SSGPE} in the chaotic regime corresponds to the onset of quantum thermalization \cite{cosme2014thermalization}.

\section{IV) Larger rings}
\begin{figure}[t]
	\begin{center}
		\includegraphics[width=	\linewidth]{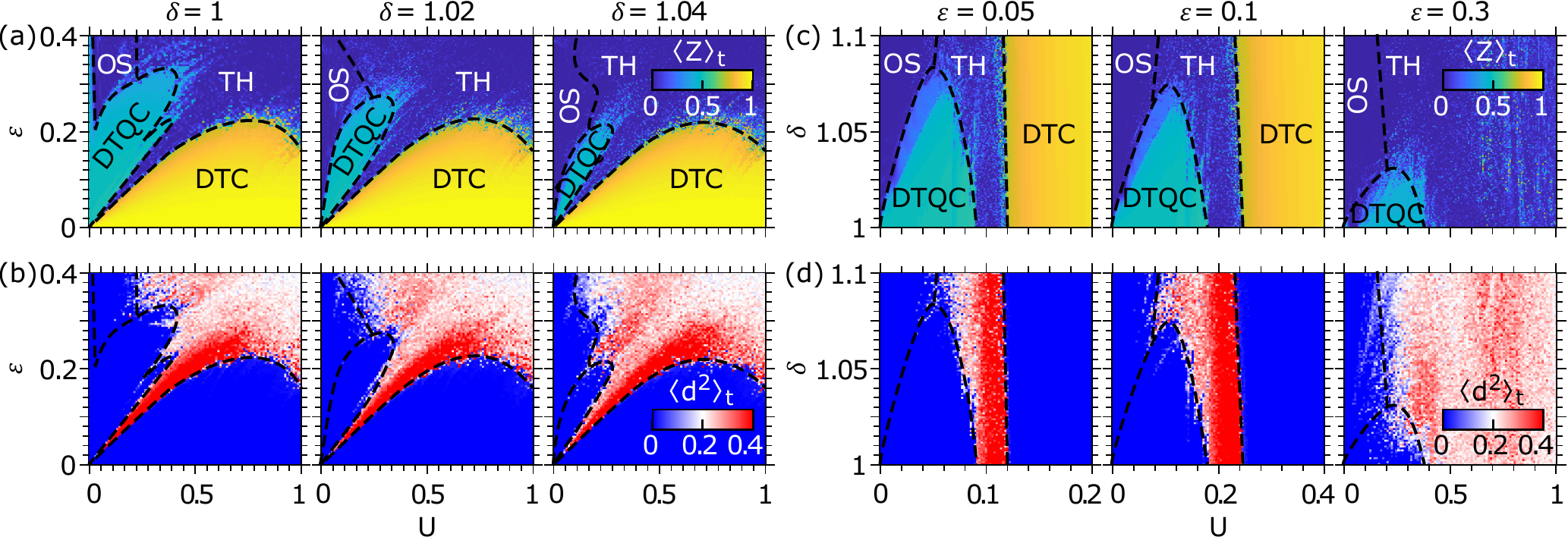}\\
	\end{center}
	\vskip -0.5cm \protect\caption
	{\textbf{Time crystalline phases in a $L = 2n = 8$-site ring}. We characterize the \gls{DTC}, \gls{DTQC}, oscillatory and thermal phases with the parameter $\langle Z \rangle_t$ (a,c) and distance $\langle d^2 \rangle_t$ between two initialy very close copies of the system (b,d), considering up to $10^4$ Floquet periods. Note that in the left panel of (a) we consider the fine-tuned limit $\delta = 1$, for which the \gls{DTQC} extends to the whole $U = 0$ axis in accordance to Eq.~(6). For $\delta \neq 1$ (central and right panel) the \gls{DTQC} is broken at $U = 0$, and possibly restored at $U \neq 0$.}
	\label{fig: L=8}
\end{figure}

In the main paper we reported numerical results for $L = 4,6$. However, our findings are valid for an arbitrary even number of sites $L =2 n \ge 4$. To support this claim, here we corroborate our results studying a system with $L = 8$ sites, for which the time crystalline phases are characterized by period $4$-tupling. To sketch the dynamical phases we look as usual at the height of the Fourier subharmonic peak $\tilde{\mathcal{I}}(\frac{2\pi}{n}) = \langle Z \rangle_t$ and at the distance $\langle d^2 \rangle_t$ between two initially very close copies of the system, which are shown in Fig.~\ref{fig: L=8}. The results that we find are completely analogue to the ones of the main text.


\end{document}

%% file: Bosons_on_ring_gls.tex

\newacronym[plural=DTCs,firstplural=discrete time crystals (DTCs)]{DTC}{DTC}{discrete time crystal}
\newacronym[plural=DTQCs,firstplural=discrete time quasicrystals (DTQCs)]{DTQC}{DTQC}{discrete time quasicrystal}

\newacronym{FC}{FC}{fully connected (or all-to-all coupled)}
\newacronym{DW}{DW}{double well}
\newacronym{ED}{ED}{exact diagonalization}
\newacronym{MF}{MF}{mean field}
\newacronym{OP}{OP}{oscillatory phase}
\newacronym{TC}{TC}{time crystal}
\newacronym{GPE}{GPE}{Gross-Pitaevskii Equation}
\newacronym{TWA}{TWA}{Truncated Wigner Approximation}
\newacronym{SSGPE}{SSGPE}{"single-shot" Gross-Pitaevskii Equation}